\documentclass[a4paper,fleqn,usenatbib]{mnras}
\usepackage{mathptmx}
\usepackage{txfonts}

\usepackage[T1]{fontenc}
\usepackage{ae,aecompl}

\usepackage{graphicx}	
\usepackage{amssymb}	
\usepackage{dcolumn}
\newcolumntype{d}[1]{D{.}{.}{#1}}
\newcommand{\sups}[1]{\textsuperscript{#1}}
\newcommand{\subs}[1]{\textsubscript{#1}}

\title[GMGPPTS]{GMRT Galactic Plane Pulsar and Transient Survey and the Discovery of PSR J1838+1523}
\author[Surnis et al.]{Mayuresh P. Surnis$^{1,2,3,4}$\thanks{E-mail: mpsurnis@ncra.tifr.res.in}, Bhal Chandra Joshi$^{1}$, Maura A. McLaughlin$^{3,4}$ \and Duncan R. Lorimer$^{3,4}$, Krishnakumar M. A.$^{2}$, P. K. Manoharan$^{2}$, Arun Naidu$^{1}$\\
$^{1}$National Centre for Radio Astrophysics, Pune, India\\
$^{2}$Radio Astronomy Centre, National Centre for Radio Astrophysics, Udhagamandalam, India\\
$^{3}$West Virginia University, Department of Physics and Astronomy, P. O. Box 6315, Morgantown, WV, USA\\
$^{4}$Center for Gravitational Waves and Cosmology, West Virginia University, Chestnut Ridge Research Building, Morgantown, WV, USA}

\begin{document}

\date{Accepted -- Received --; in original form --}

\pagerange{\pageref{firstpage}--\pageref{lastpage}} \pubyear{0000}

\maketitle

\label{firstpage}

\begin{abstract}
We report the results of a blind pulsar survey carried out with the Giant Metrewave Radio Telescope (GMRT) at 325 MHz. The survey covered about 10\% of the region between Galactic longitude 45$^{\circ} < l <$ 135$^{\circ}$ and Galactic latitude 1$^{\circ} < |b| <$ 10$^{\circ}$ with a dwell time of 1800 s, resulting in the detection of 28 pulsars. One of these, PSR J1838+1523, was previously unknown and has a period of 549 ms and a dispersion measure of 68 pc cm\sups{$-$3}. We also present the timing solution of this pulsar obtained from multi-frequency timing observations carried out with the GMRT and the Ooty Radio Telescope. The measured flux density of this pulsar is 4.3$\pm$1.8 and 1.2$\pm$0.7 mJy at 325 and 610 MHz, respectively. This implies a spectral index of $-$2$\pm$0.8, thus making the expected flux density at 1.4 GHz to be about 0.2 mJy, which would be just detectable in the high frequency pulsar surveys like the Northern High Time Resolution Universe pulsar survey. This discovery underlines the importance of low frequency pulsar surveys in detecting steep spectrum pulsars, thus providing complementary coverage of the pulsar population.   
\end{abstract}

\begin{keywords}
(stars:) pulsars: general -- (stars:) pulsars: individual (PSR J1838+1523) -- surveys: astronomical databases
\end{keywords}

\section{Introduction}
\label{intro}
The active pulsar population in our Galaxy has been estimated to be about a million \citep{fk06}, while the number of currently known pulsars is only about 2600\footnote{http://www.atnf.csiro.au/people/pulsar/psrcat/}. On top of that, the observed pulsar emission shows a lot of temporal variations that range from nulls \citep{b70}, lasting a few periods, and extreme ON-OFF states with durations of days to months, shown by intermittent pulsars \citep{klo+06,crc+12,llm+12}, to transient isolated bursts over time scales of hours, shown by rotating radio transients \citep[RRATs;][]{mll+06}. This makes repeated pulsar searches of the same area of sky more fruitful in discovering these interesting objects.

The majority of past and current pulsar surveys have used large single dish telescopes and $\sim$1 GHz observing frequencies with shorter integration times as a trade off between sky background (which reduces the sensitivity at low frequencies) and survey speed (slow at high frequencies). A multi-element telescope like the Giant Metrewave Radio Telescope (GMRT) provides large collecting area of the order of 30,000 m\sups{2} with a field of view of about 1\sups{$\circ$} at 325 MHz \citep{sak+91}. This makes it a sensitive, rapid pulsar search instrument. In addition, a pulsar survey carried out with the GMRT at 325 MHz would prove to be complementary to high frequency pulsar surveys in providing a complete coverage of pulsar parameters by discovering very steep spectrum pulsars.

With this motivation, \cite{jml+09} carried out a blind pulsar survey with the GMRT at 610 MHz. This survey covered a region between Galactic longitude 45$^{\circ} < l <$ 135$^{\circ}$ and Galactic latitude $|b| <$ 1$^{\circ}$ and resulted in the discovery of three pulsars. In this work, we present a follow-up survey done at higher Galactic latitudes at a frequency of 325 MHz. The survey, named GMRT Galactic Plane Pulsar and Transient Survey (GMGPPTS), covered about 10\% of the region between Galactic longitude 45$^{\circ} < l <$ 135$^{\circ}$ and Galactic latitude 1$^{\circ} < |b| <$ 10$^{\circ}$. A total of 28 pulsars were detected in this survey including one new discovery. Here, we also report the timing solution of the newly discovered pulsar, PSR J1838+1523, obtained through multi-frequency timing observations carried out with the GMRT and the Ooty Radio Telescope (ORT).

In Section \ref{obs}, we describe the observational set-up. Section \ref{anal} outlines the search data analysis. Sensitivity of the survey is discussed in Section \ref{sens}. The survey results are mentioned in Section \ref{res}. The discovery and follow-up of the newly discovered PSR J1838+1523 are described in Section \ref{1838}. We discuss the implications of our results for future low frequency surveys in Section \ref{disc}. Finally, we present our conclusions in Section \ref{concl}. 

\section{Survey Observations}
\label{obs}
\begin{figure}
\begin{center}

\includegraphics[scale=0.6]{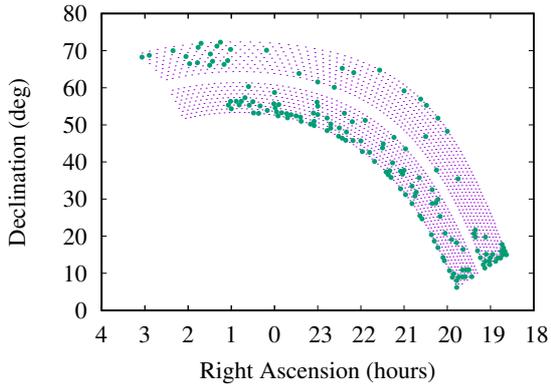}
\caption{The region of sky observed in GMGPPTS. Right ascension is plotted in hours and declination in degrees. Each small dot represents the phase center of an individual pointing. Filled cyan circles indicate the observed fields. The blank area in between is the Galactic plane.}
\label{obs_point}

\end{center}
\end{figure}

We carried out the survey observations with the hardware back-end of the GMRT. This back-end combined powers from each of the 30 GMRT dishes to form the incoherent array (IA) output. We acquired the data over a bandwidth of 16 MHz. The hardware back-end provided 256 spectral channels across the band-pass with a sampling time of 256~$\mu$s in 16-bit float format. The survey area was divided into 1548 individual pointings (circular region of diameter $\sim$1$^{\circ}$) and we observed each pointing for 1800 s. The observations were carried out in two observing cycles of the GMRT. In each cycle, we divided the time into 5 sessions of 10 hours each, placed on consecutive days. We completed the first set of observations between 2009 July 23--27 and the second set of observations between 2009 December 25--29. We observed a total of 152 pointings, covering an area of about 115 deg\sups{2}. Figure \ref{obs_point} shows the pointings (purple dots) covering the proposed survey region as well as the observed pointings (filled cyan circles). The raw data were written to magnetic tapes as well as hard disks for further analysis. We manually prepared detailed observing logs for each sub-session and carefully backed up all the data and the auxiliary files containing time stamps.

\section{Data Analysis}
\label{anal}

\subsection{Pre-processing}
\label{anal_pre}
We started the data analysis with the removal of radio frequency interference (RFI). A program developed by us was used to produce root-mean-square (rms) for each spectral channel after a 1 s running mean subtraction over the full 1800 s integration. We masked the channels at the beginning and the end of the bandpass (typically 10 channels from each side). The channels affected by narrow-band RFI usually showed significantly higher rms than the neighbouring channels. Hence, we also masked the channels with rms larger than twice the typical rms in the middle 80\% band after a manual inspection of a plot of rms as a function of spectral channel. This flagging resulted in an effective bandwidth of 14 MHz. In addition, we also removed a 10 s running mean from each of the unmasked channels to remove baseline variations. We then removed broad-band RFI by subtracting the mean of all spectral channels from each time sample, the so called zero dispersion measure (DM) filtering technique \citep{ekl09} and converted the raw data into 16-bit filterbank format. 

\subsection{Pulsar Search}
\label{anal_srch}
We initially analysed these data on a high performance cluster having 32 compute nodes, each with a dual-core processor. The analysis was done using a search pipeline based on \textsc{sigproc}\footnote{http://www.sigproc.sourceforge.net}. The search was parallelised using \textsc{open-mpi}\footnote{http://www.open-mpi.org} with the embarrassingly parallel model wherein, for a given pointing, each node carried out the pulsar search independently on its own list of trial DMs. This pipeline searched for periodicity by summing up to 32 harmonics (five harmonic sums) with a threshold of eight sigma. We also carried out single-pulse search using the method described by \cite{cm03} at each trial DM up to a maximum smoothing of 262 ms with a threshold of six sigma. 

We re-analysed these data on a new 10 teraflops cluster at NCRA using a search pipeline based on \textsc{presto}\footnote{http://www.cv.nrao.edu/$\sim$sransom/presto}. The \textsc{presto} based pipeline used sub-band de-dispersion to reduce computational demand \citep{mks+11} and hence, could produce the full range of de-dispersed time-series files on a single node. We used 256 sub-bands for de-dispersion. This resulted in maximum smearing of 1.5, 8 and 20 ms at DMs of 100, 500 and 1000~pc~cm$^{-3}$, respectively. We used \textsc{presto} to carry out an acceleration search, which is necessary for discovering short orbital period binary pulsars. We used a \textsc{zmax} of 200 in \textsc{accelsearch}, corresponding to an acceleration of 18.5 m s\sups{$-$2} for a pulsar with a spin period of 1 ms \citep[see][for details]{rem02}. We did not use \textsc{presto} for single-pulse search. We chose the trial DM range of 0--1000 pc cm\sups{$-$3} for the pulsar search. We divided this range in 6 sub-ranges. We searched the range 0--57.6 pc cm\sups{$-$3} with a DM step of 0.1 pc cm\sups{$-$3}, 57.6--112.2 pc cm\sups{$-$3} with a DM step of 0.3 pc cm\sups{$-$3}, 112.2--197.7 pc cm\sups{$-$3} with DM step of 0.5 pc cm\sups{$-$3}, 197.7--390.7 pc cm\sups{$-$3} with a DM step of 1 pc cm\sups{$-$3}, 390.7--778.7 pc cm\sups{$-$3} with a DM step of 2 pc cm\sups{$-$3} and 778.7--1000 pc cm\sups{$-$3} with a DM step of 5 pc cm\sups{$-$3}, calculated using the \textsc{presto} tool \textsc{DDplan.py}. We did not search at DMs larger than 1000 pc cm\sups{$-$3} as the radio pulses (from pulsars or fast radio bursts) would be scatter-broadened by more than a few milliseconds at 325 MHz even at the highest Galactic latitude that we observed. 

\subsection{Candidate Selection}
\label{anal_candsel}
\begin{figure}
\begin{center}

\includegraphics[trim=0 0 0 3cm, clip, scale=0.45]{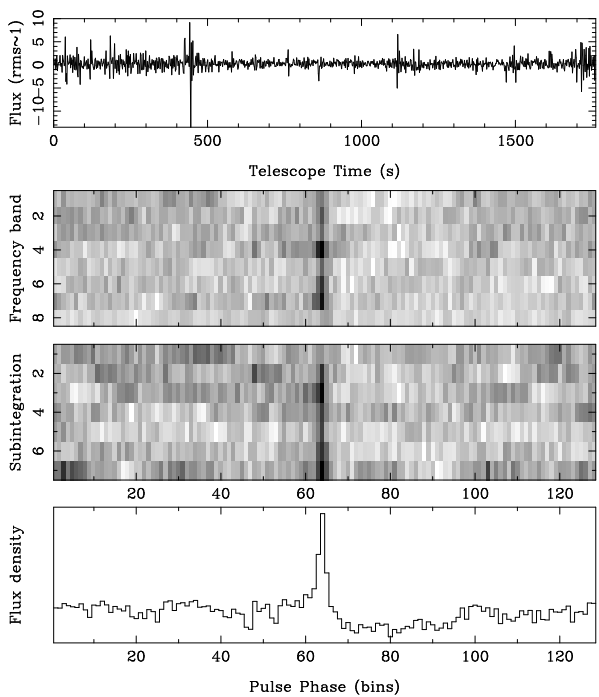}
\caption{The diagnostic candidate plot showing the discovery of PSR J1838+1523. From top to bottom, the first panel shows de-dispersed, zero mean time-series, the second panel shows grey scale plot of folded profile in 8 frequency sub-bands, the third panel shows grey scale plot of folded profile in 8 time sub-integrations and the fourth panel shows the folded profile for the full bandwidth and time.}
\label{1838_disc}

\end{center}
\end{figure}

We manually scrutinized the candidate plots generated by the \textsc{sigproc} based pipeline. One such plot is shown in Figure \ref{1838_disc} showing the discovery of PSR J1838+1523. For the \textsc{presto} based pipeline, we used the built-in candidate sifting script \textsc{accel\_sift.py} to prune the list of candidates. In the sifting, we rejected the candidates below a significance of eight sigma. In addition, we also rejected candidates with periods below 0.5 ms or above 15 s  as well as candidates which appeared at only one trial DM. We produced the diagnostic plots only for the selected candidates and scrutinized them manually. 

\section{Survey Sensitivity}
\label{sens}
We calculated the overall survey sensitivity using the radiometer equation, which gives the minimum detectable flux density 
\begin{equation}
S_{{\rm min}}  =  \frac{S/N_{{\rm min}}T_{{\rm sys}}}{G \sqrt{n_{A}n_{p}t_{{\rm obs}}\triangle f}} \sqrt{\frac{W}{P - W}}
\label{sens_expsnr}
\end{equation}

We assumed a system temperature ($T_{{\rm sys}}$) of 190 K [receiver temperature of 66 K \citep{l13} and sky background of 124 K (averaged over the survey region), scaled to 325 MHz from the \cite{hss+82} sky map], telescope gain ($G$) of 0.32 K Jy\sups{$-$1} Antenna\sups{$-$1} for 25 GMRT antennae ($n_A$) with 2 polarizations summed ($n_p$), and a cut-off S/N of 8 ($S/N_{{\rm min}}$). In the above equation, $P$ and $W$ are the pulse period and width of a top-hat pulse with equal area. We assumed a pulse width of 10\% of the period for our calculations. In order to get a practical estimate of the sensitivity, the theoretical sensitivity was multiplied by a degradation factor, which was given by the mean ratio of expected to observed profile S/N for known pulsars. The expected S/N were calculated using the above stated assumptions and the radiometer equation with $S_{\rm min}$ replaced by $S_{\rm mean}$, which is the pulse averaged flux density of the pulsar in mJy. We scaled the mean flux densities of the known pulsars to the observing frequency using the available spectral indices from the pulsar catalogue\footnote{http://www.atnf.csiro.au/people/pulsar/psrcat/} \citep{mht+05}. In cases, where spectral index was not available, we assumed it to be $-$1.4 \citep{blv13}. We further corrected the frequency scaled flux densities for the beam response at 325 MHz given the offset in the position of the pulsar with respect to the beam center. We estimated pulse widths for detected pulsars from their observed profiles. In the case of undetected pulsars, we used their pulse widths (w\subs{10}) given in the pulsar catalogue. We assumed a pulse width of 10\% whenever published pulse widths were not available.

In order to estimate the sensitivity for the single-pulse search, we produced pulse energy sequences in the ON and OFF pulse regions for known pulsars. The de-dispersed time-series was folded modulo the pulsar period to produce a phase-time plot showing each pulse period. ON and OFF pulse widows of equal width were then manually chosen with the help of an interactive program to produce sequences of pulse energies as separate files. We then calculated the observed single pulse S/N by taking a ratio of the mean of ON pulse energies to the rms of the OFF pulse energies in an RFI-free region. We calculated the expected single pulse S/N using the radiometer equation

\begin{equation}
S/N  =  \frac{S_{{\rm mean}}G \sqrt{n_{A}n_{p}t_{{\rm obs,sp}}\triangle f}}{T_{{\rm sys}}}
\label{sens_sp}
\end{equation}

\noindent with $t_{{\rm obs,sp}}$ equal to the duration of the ON (or OFF) pulse window. The ratio of the expected to observed S/N provided the degradation factor for scaling the theoretical sensitivity to get a practical estimate of the survey sensitivity. To estimate the degradation factor, we used only those pulsars where observed S/N did not differ from the expected S/N by more than a factor of five to exclude cases of large amounts of RFI affecting the sensitivity. We report the  degradation factors for the harmonic and single-pulse search in Section \ref{res}.

\section{Results}
\label{res}
\begin{figure}
\begin{center}

\includegraphics[angle=-90,scale=0.35]{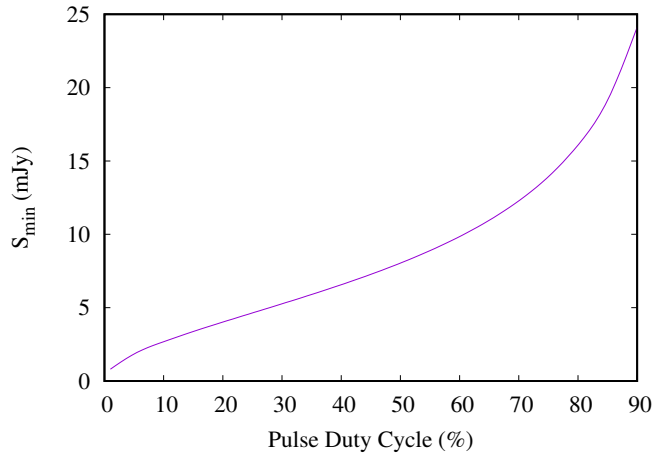}
\caption{The overall survey sensitivity ($S_{min}$) as a function of the pulse duty cycle for a period of 10 ms and a DM of 100 pc cm\sups{$-$3}.}
\label{res_sens}

\end{center}
\end{figure}

We detected a total of 28 pulsars out of which one, PSR J1838+1523 (see Figure \ref{1838_disc}), is a newly discovered pulsar. The results of the follow-up timing observations for this pulsar are discussed in detail in the next section. The observed fields contained 32 known pulsars. Out of the five non-detections, the expected scintillation time scale for PSR B1907+12 is about 3000 s, much longer than the integration time of 1800 s. PSR J0103+54 has no measured flux density and a position uncertainty of the order of 0.1$^{\circ}$ \citep{kkl+15} hence, it could be on the edge of the GMRT IA beam. The other three pulsars are new discoveries \citep{slr+14} and have a large position uncertainty of the order of 0.8$^{\circ}$. Table \ref{res_knpsrtab} lists the detected known pulsars with their parameters. The table also lists expected and observed profile S/N with appropriate remarks for discrepancies and non-detections. One point worth mentioning with respect to the observed DMs is that for some pulsars like PSRs B0105+65, B1919+20, B1919+21, B1920+21 and B1929+10, they are different than the catalogued DMs (see Table \ref{res_knpsrtab}). While the catalogued DMs are usually estimated from multi-frequency timing analysis, the DMs estimated from search analysis can be offset from true pulsar DM. One glance at Table \ref{res_knpsrtab} reveals that wherever the search DMs are offset, the pulse was distorted due to zero DM filtering. This could be a dominant reason for the search DMs to be slightly offset, especially if the pulsar is strong. We also detected 13 out of the 32 known pulsars in the single-pulse search. Although we did not search at very large DMs, we did not detect any fast radio bursts in the single-pulse search. We estimated the expected theoretical sensitivity for an eight sigma detection assuming a 10\% pulse duty cycle to be 1.4 mJy. Excluding the pulsars which showed either very low or very high observed S/N as compared to expected S/N (either due to RFI or scintillation), the mean degradation factor was found to be 1.9. The major source of degradation in the sensitivity is residual RFI present in the data. Thus, the practical eight sigma detection limit for a pulsar with a spin period of 10 ms and a DM of 100 pc cm\sups{$-$3}, having 10\% duty cycle for our survey is 2.7 mJy. The detection limits for other duty cycles can be inferred from the sensitivity curve plotted in Figure \ref{res_sens}. We estimated the expected eight sigma minimum detectable flux density for the single-pulse search (assuming a pulse width of 10 ms) to be 1.8 Jy, while the mean observed degradation factor was found to be 1.3. This implies a corrected value of 2.3 Jy or a fluence of 23 mJy s.

\begin{table*}
\begin{minipage}{175mm}
\begin{scriptsize}

\caption{Parameters for known pulsars in the region covered by GMGPPTS. The columns from left to right are, pulsar name, observed period, catalogue period, observed DM, catalogue DM, expected S/N, observed S/N, the distance of the pulsar from the field center, whether single pulses were detected, and remarks for any discrepancies. The catalogue period and DM as well as parameters for all un-detected pulsars were taken from the ATNF pulsar catalogue (version 1.54).}
\label{res_knpsrtab}

\begin{tabular}{cd{4.1}d{4.1}d{3.1}d{3.1}d{4.0}d{4.0}ccc}

\hline
\multicolumn{1}{c}{Name} & \multicolumn{1}{c}{P\subs{obs}} & \multicolumn{1}{c}{P\subs{cat}} & \multicolumn{1}{c}{DM\subs{obs}}   & \multicolumn{1}{c}{DM\subs{cat}}   & \multicolumn{1}{c}{Expected} & \multicolumn{1}{c}{Observed} & Distance & Single Pulses & Remarks \\
                         & \multicolumn{1}{c}{(ms)}        & \multicolumn{1}{c}{(ms)}        & \multicolumn{1}{c}{pc cm\sups{-3}} & \multicolumn{1}{c}{pc cm\sups{-3}} & \multicolumn{1}{c}{S/N}                        & \multicolumn{1}{c}{S/N}      & (\sups{$\circ$}) & Detected ? &            \\
\hline

J0103+54   &      * &  354.3 &     * &  55.6 &    * &    * & 0.5 & N & Large position uncertainty \\
B0105+65   & 1283.7 & 1283.7 &  32.5 &  30.5 &  189 &  116 & 0.2 & Y & Pulse distortion due to zero DM filtering \\
J0112+66	   &      * & 4301.2 &     * & 112.0 &   98 &    * & 0.2 & N & Large position uncertainty \\
B1842+14   &  375.5 &  375.5 &  41.0 &  41.5 &  115 &   45 & 0.9 & N & \\
B1848+12   & 1205.3 & 1205.3 &  71.6 &  70.6 &  110 &   25 & 0.6 & N & Pulse distortion due to zero DM filtering \\
B1848+13   &  345.6 &  345.6 &  61.1 &  60.1 &   54 &   27 & 0.6 & N & \\
J1853+1303 &    4.1 &    4.1 &  30.6 &  30.6 &   62 &   26 & 0.2 & N & \\
B1907+10   &  283.6 &  283.6 & 150.4 & 150.0 &  724 &  422 & 0.5 & Y & \\
B1907+12   &      * & 1441.7 &     * & 258.6 &   45 &    * & 0.2 & N & Scintillation time of 3000 s at 325 MHz \\
B1911+13   &  521.5 &  521.5 & 144.7 & 145.1 &   70 &   37 & 0.3 & N & \\
B1915+13   &  194.6 &  194.6 &  94.7 &  94.5 &  222 &  159 & 0.9 & N & \\
B1918+19   &  821.0 &  821.0 & 153.2 & 153.9 &  387 &  102 & 0.2 & N & \\
B1919+20   &  380.3 &  760.7 &  97.4 & 101.0 &   31 &   15 & 0.3 & N & \\
B1919+21   &  668.7 & 1337.3 &   8.2 &  12.4 & 1210 &  403 & 0.3 & Y & Pulse distortion due to zero DM filtering \\
B1920+21   & 1078.0 & 1078.0 & 214.7 & 217.1 &  415 &  330 & 0.6 & Y & \\
B1929+10   &  226.5 &  226.5 &   6.7 &   3.4 & 2635 &  578 & 0.7 & Y & Pulse distortion due to zero DM filtering \\
B1933+16   &  358.7 &  358.7 & 158.6 & 158.5 & 2360 & 1376 & 0.6 & Y & \\
J1944+0907 &    5.2 &    5.2 &  24.4 &  24.3 &   12 &   15 & 0.2 & N & \\
B1944+17   &  440.6 &  440.6 &  17.3 &  16.2 &  446 &   14 & 0.2 & N & Baseline affected by 50 Hz harmonic (440 ms) \\
B1946+35   &  717.3 &  717.3 & 129.2 & 129.1 &  692 &  595 & 0.7 & Y & \\
B2016+28   &  557.9 &  557.9 &  15.3 &  14.2 & 3066 &  672 & 0.7 & Y & Pulse distortion due to zero DM filtering \\
B2020+28   &  343.4 &  343.4 &  27.0 &  24.6 &  701 &  384 & 0.3 & Y & \\
J2030+55   &  289.5 &  579.0 &  60.5 &  60.0 &    * &   37 & 0.3 & N & No flux measurements available \\
B2035+36   &  618.7 &  618.7 &  91.7 &  93.6 &   81 &   34 & 0.7 & N & \\
J2102+38   & 1190.0 & 1189.9 &  85.7 &  85.0 &    * &   26 & 0.4 & N & No flux measurements available \\
B2111+46   & 1014.6 & 1014.7 & 144.2 & 141.3 &  609 &  893 & 0.3 & Y & \\
J2137+64   &      * & 1750.9 &     * & 106.0 &   35 &    * & 0.5 & N & Large position uncertainty \\
B2217+47   &  538.4 &  538.5 &  46.4 &  43.5 & 1030 & 2446 & 0.2 & Y & \\
B2224+65   &  682.5 &  682.5 &  38.0 &  36.1 &  478 &   26 & 0.4 & N & Large amount of unaccounted RFI \\
J2229+64   &      * & 1893.1 &     * & 194.0 &   21 &    * & 0.4 & N & Large position uncertainty \\
J2238+6021 & 1023.4 & 3070.2 & 185.7 & 182.0 & 2517 &   42 & 0.2 & Y & Large position uncertainty \\
B2323+63   & 1436.2 & 1436.3 & 199.7 & 197.4 &  190 &   41 & 0.6 & Y & Pulse distortion due to zero DM filtering \\ 

\hline

\end{tabular}
\end{scriptsize}
\end{minipage}
\end{table*}

\begin{figure}
\begin{center}

\includegraphics[scale=0.65]{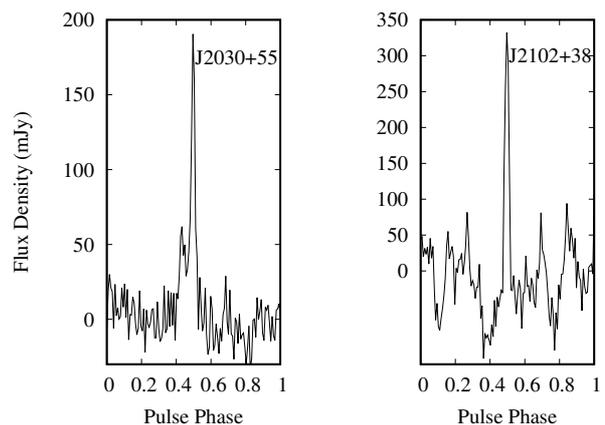}
\caption{Pulse profiles for PSRs J2030+55 (left) and J2102+38 (right) at a frequency of 325 MHz.}
\label{res_knpsr}

\end{center}
\end{figure}

We are also reporting pulse averaged flux densities at 325 MHz for PSRs J2030+55 and J2102+38 for the first time from the observed profiles (Figure \ref{res_knpsr}) using the radiometer equation (see Equation \ref{sens_expsnr}). The estimated pulse averaged flux density for PSR J2030+55 is 7.2$\pm$0.6 mJy and for PSR J2102+38, it is 4.4$\pm$0.5 mJy. The errors in flux density are the three sigma radiometer noise values for the same parameters as given in Section \ref{sens}.

\section{PSR J1838+1523}
\label{1838}

\subsection{Initial Timing Observations at the GMRT}
\label{1838_initfollow}
Immediately after the confirmation as a pulsar, we carried out gridding observations [as described by \citep{jml+09}] for PSR J1838+1523 to reduce the position uncertainty. We followed this with regular timing observations between 2011 February 12 $-$ 2014 March 29 (MJD 55604$-$56745) at a cadence of two weeks at 325 MHz with the GMRT in the IA mode using the GMRT software back-end \citep[GSB;][]{rgp+10}. This back-end provided 512 spectral channels across a bandwidth of 33 MHz with 122.88 $\mu$s sampling. We recorded interferometry data simultaneously with a sampling time of 16 s. We used an integration time of 1800 s in each session. For interferometry observations, we observed 3C48 and 3C286 as flux calibrators and 1822$-$096 and 1859+129 as phase calibrators. During the initial gridding observations (wherein the position uncertainty was still 0.6$^{\circ}$), we noticed that the pulsar was not detected in all the observations carried out between 2011 August 30 (MJD 55803) and 2012 June 13 (MJD 56090). After this, the pulsar showed a random set of detections and non-detections on a time scale of one--two months. When the pulsar was detected again in observations carried out on 2013 December 27 (MJD 56653), we decided to do gridding observations with the ORT. 

\subsection{Timing Observations at the ORT}
\label{1838_ortfollow}
ORT has a fan shaped beam with a beam width of about 2.3$^{\circ}$ in right ascension and 6$'$ in declination. We used this fact to construct an array of 15 pointings along the declinations between +14$^{\circ}$25$'$ to +15$^{\circ}$49$'$ with a separation of 6$'$. We carried out the gridding observations on 2013 December 31 (MJD 56657) wherein, we observed each pointing for 15 minutes. The pulsar was detected in two of the 15 pointings. We then observed three more declination positions separated by 3$'$ centred at the declination of +15$^{\circ}$25$'$ for 15 minutes each. Based on the observed S/N, we narrowed down the declination position of the pulsar to +15$^{\circ}$22$'$. We immediately started daily follow-up observations at the ORT. These observations continued until 2015 July 31 (MJD 57234). The cadence was decreased to twice per week from then onwards. We used the Pulsar Ooty Radio Telescope New Digital Efficient Receiver \citep[PONDER;][]{njm+15}, which provided coherently de-dispersed time-series data over 16 MHz bandwidth with 500 $\mu$s sampling for an integration time of 1800 s for each epoch. We also observed 3C93 and 3C94 as flux density calibrators.   

\subsection{Multi-frequency Timing Observations at the GMRT}
\label{1838_multifollow}
We obtained a phase connected timing solution from ORT timing observations spread over a six month period. We confirmed the refined pulsar position thus obtained, using an observing session at 820 MHz with the Green Bank Telescope (GBT) on 2014 September 11 in search mode with a bandwidth of 200 MHz and a sampling time of 82 $\mu$s. The data were recorded for 1800 s with the Green Bank Ultimate Pulsar Processing Instrument \citep[GUPPI;][]{drd+08}. Once the pulsar position was identified with good accuracy, we observed the pulsar at 1170 MHz (2 epochs) and 610 MHz (11 epochs) with the GMRT, with simultaneous ORT observations at 325 MHz, in order to determine a more precise DM and spectral index of the pulsar. We carried out these observations between 2014 October 25 (MJD 56955) and 2015 September 17 (MJD 57282) recording simultaneous phased array and interferometry data with a cadence of about a month.

\subsection{Time-series Analysis and Results}
\label{timing}
\begin{figure}
\begin{center}

\includegraphics[scale=0.65]{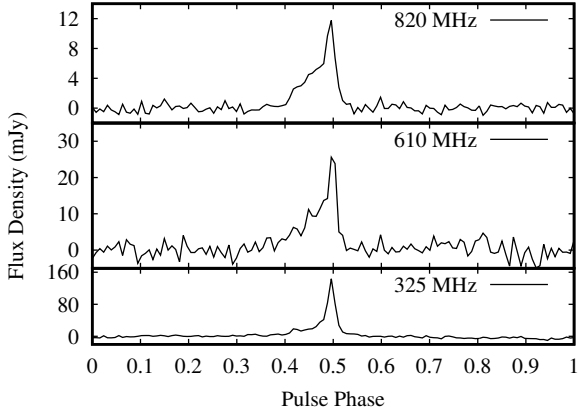}
\caption{Average profiles of PSR J1838+1523 at different frequencies. The profiles at 325 and 610 MHz were scaled using interferometric measurements, while the 820 MHz GBT profile was scaled using extrapolated flux density assuming a spectral index of $-$2. The effective integration times for the profiles are 0.5, 5 and 20 hours at 820, 610 and 325 MHz, respectively.}
\label{1838_prof}

\end{center}
\end{figure}

Both of the observations at 1170 MHz with the GMRT resulted in non-detections, while at 610 MHz, there were 4 non-detections. On the other 7 epochs, we obtained folded profiles from GMRT as well as ORT data using \textsc{sigproc}. We used flux density calibrator scans on each epoch to calibrate these profiles. The resultant profiles at different observing frequencies are shown in Figure \ref{1838_prof}.

\begin{figure}
\begin{center}

\includegraphics[scale=0.65]{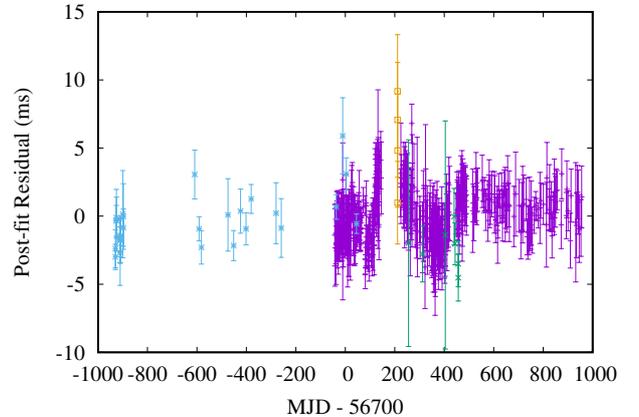}
\caption{Post-fit timing residuals obtained for PSR J1838+1523. The rms post-fit residual is 1.6 ms. The blue asterisks indicate TOAs measured at 325 MHz, while green crosses indicate TOAs measured at 610 MHz with the GMRT. Purple crosses indicate TOAs measured with the ORT at 326 MHz and yellow squares indicate TOAs measured with the GBT at 820 MHz. The gap in ORT TOAs is due to the telescope being down for maintenance.}
\label{1838_timres}

\end{center}
\end{figure}

At each observing frequency, we added the observed profiles after matching their peaks to form high S/N (S/N $\sim$ 200) template profiles. We determined the pulse times-of-arrival (TOAs) using a standard template matching technique as described by \cite{t92}. We then carried out the timing analysis using \textsc{tempo2}\footnote{www.atnf.csiro.au/research/pulsar/tempo2}\citep{hem06}. During the timing analysis, we noticed that the TOA uncertainties reported by the cross-correlation analysis were underestimated. We thus used a multiplicative factor (EFAC) to get final TOA uncertainties. We determined the EFACs for different telescopes and frequencies independently such that the reduced $\chi^2$ value of the timing model fit was 1. The resultant EFACs were 2.5 and 5.3 for GMRT TOAs at 325 and 610 MHz, respectively. The EFACs for ORT and GBT TOAs were found to be 2.7 and 1.6, respectively. Table \ref{1838_timsol} shows the timing solution obtained from multi-frequency timing observations while the timing residuals are shown in Figure \ref{1838_timres}. It should be noted that the DM was also determined from multi-frequency TOAs.

\begin{table}

\begin{tabular}{ll}
\hline \hline
\multicolumn{2}{c}{Fit and data-set} \\
\hline
Pulsar name\dotfill                   & J1838+1523 \\ 
MJD range\dotfill                     & 55770--57655 \\ 
Data span (yr)\dotfill                & 5.2 \\ 
Number of TOAs\dotfill                & 386 \\
RMS timing residual (ms)\dotfill      & 1.6 \\ 
\hline
\multicolumn{2}{c}{Measured Quantities} \\ 
\hline
Pulse frequency, $\nu$ (s$^{-1}$)\dotfill                           & 1.820960929021(4) \\ 
First derivative of pulse frequency, $\dot{\nu}$ (s$^{-2}$)\dotfill & $-$6.724(2)$\times 10^{-17}$ \\ 
Dispersion measure, DM (pc cm$^{-3}$)\dotfill                       & 68.26(3) \\
\hline
\multicolumn{2}{c}{Set Quantities} \\ 
\hline
Epoch (MJD)\dotfill          & 56675 \\ 
\hline
\multicolumn{2}{c}{Derived Quantities} \\
\hline
Right ascension, $\alpha$ (hh:mm:ss)\dotfill                        &  18:38:46.78(1) \\ 
Declination, $\delta$ (dd:mm:ss)\dotfill                            & +15:23:25.4(1) \\
Galactic longitude, l ($^{\circ}$)\dotfill                          & 45.34730(4) \\
Galactic latitude, b ($^{\circ}$)\dotfill                           & +9.69399(4) \\
$\log_{10}$(Characteristic age, yr) \dotfill             &  8.6 \\
$\log_{10}$(Surface magnetic field strength, G) \dotfill & 11.0 \\
$\log_{10}$(\.{E}, ergs/s) \dotfill                      & 30.7 \\
\hline
\multicolumn{2}{c}{Assumptions} \\
\hline
Clock correction procedure\dotfill   & TT(TAI) \\
Solar system ephemeris model\dotfill & DE405 \\
Binary model\dotfill                 & NONE \\
Model version number\dotfill         & 5.00 \\ 
\hline \hline
\end{tabular}
\caption{Timing solution obtained for PSR J1838+1523. (Note: Figures in parentheses are the nominal one sigma \textsc{tempo2} uncertainties in the least-significant digits quoted.)}
\label{1838_timsol}
\end{table}

\begin{figure}
\begin{center}

\includegraphics[scale=0.65]{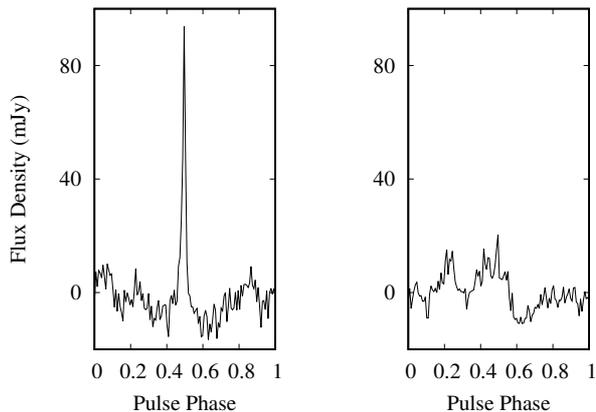}
\caption{Left Panel: Average profile obtained after adding the individual profiles obtained from the 15 weak detections corresponding to an integration time of 7.5 hours. Right Panel: Average profile obtained after adding all the non-detections corresponding to an integration time of 4.5 hours.}
\label{weakoffprof}

\end{center}
\end{figure}
When we used the final timing solution to fold the initial non-detections with the GMRT at 325 MHz, 15 out of 30 epochs resulted in weak detections with the profile S/N lying between 8--10. The average profile obtained after averaging these weak detections is shown in Figure \ref{weakoffprof} along with average profile for all non-detections.

\subsection{Imaging Analysis and Results}
\label{imaging}
\begin{figure}
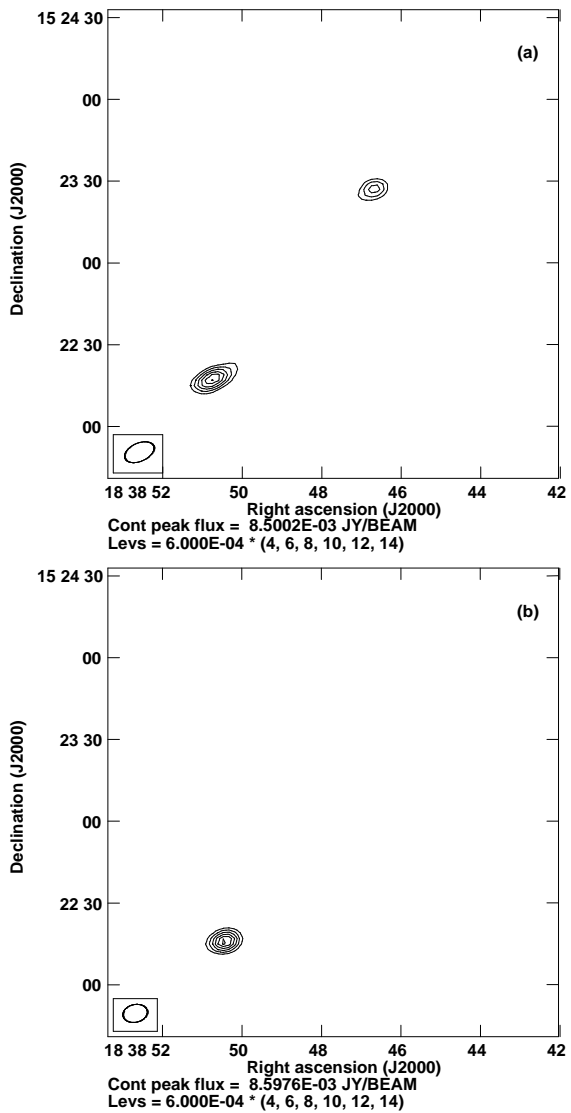


\includegraphics[trim=0 0 0 1.4cm, clip, scale=0.4]{J1838+1523_on.eps}
\includegraphics[trim=0 0 0 1.4cm, clip, scale=0.4]{J1838+1523_off.eps}
\caption{ Interferometric images made at 325 MHz showing radio source NVSS J183850+152214 (towards bottom left) and PSR J1838+1523 (towards top right) at two epochs. {\it (a):} The image made on 2012 December 1 (MJD 56262). The pulsar was detected in the time-series on this epoch. {\it (b):} The image made on 2013 March 30 (MJD 56381). The pulsar was not detected in the time series on this epoch. In both panels, the image rms is 0.6 mJy. The contours are at 2.4, 3.6, 4.8, 6.0, 7.2 and 8.4 mJy. The synthesized beam is plotted in an inset on the bottom left corner.}
\label{1838_imag}

\end{figure}

Given the intermittent nature of detections for this pulsar and consequent difficulty in getting a good timing solution, we resorted to making images from the interferometry data to obtain a more accurate position. To avoid confusion with other nearby radio point sources, we used the following logic. If the pulsar is detected in the time-series, it will be present as a continuum source in the image and not otherwise. In order to make images, the data had to be flagged and calibrated carefully. We used a program \textsc{flagcal}\footnote{http://hdl.handle.net/2301/581}\citep{c13} for this purpose. We then made images with the Astronomical Image Processing System (AIPS)\footnote{http://www.aips.nrao.edu/index.shtml} package. We used multi-facet imaging technique with 19 facets spread across the primary beam to correct for the W-projection effect \citep{c99}. We also carried out 3 rounds of self-calibration with solution intervals of 4, 1, and 0.5 minutes to get the final image at each epoch. The detection in the PA mode observations with the GMRT at 610 MHz helped in narrowing down the region (to less than 100$''$) to look for the continuum counterpart of the pulsar. This region contained only two continuum sources in the images at 325 MHz. One of these sources was identified as NVSS 183850+152214 by comparing with the NRAO VLA Sky Survey \citep[NVSS;][]{ccg+98} image at 1400 MHz. The other source, located at RA 18$^h$ 38$^m$ 47$^s$, DEC +15$^{\circ}$ 23$'$ 27$''$ with an error of about 6$''$, did not have a counterpart in the NVSS map, implying a steep spectrum. In addition, NVSS 183850+152214 turned out to be a very flat spectrum source with flux densities of 8.3$\pm$1.5 and 7.0$\pm$0.5 mJy at 325 and 1400 MHz, respectively. This indicated that the unidentified source could be the pulsar. The radio image of this region for one epoch is shown in panel (a) in Figure \ref{1838_imag}. This conjecture was confirmed when the steep spectrum source was not detected in an image made on one epoch which resulted in non-detection of the pulsar in the time series as well (upper limit on continuum flux density from the radio map is consistent with the expected detection threshold calculated from Equation \ref{sens_expsnr}). Panel (b) in Figure \ref{1838_imag} shows the radio image of this epoch. The position of the pulsar obtained in the final timing solution (see Table \ref{1838_timsol}) is consistent with the position of the continuum source, thus providing an independent confirmation. When we made images for other epochs with time-series non-detections, we found that the continuum source was present. This supported the possibility of the pulsar having un-pulsed emission. We discuss the implications of this conjecture in Section \ref{disc}.

\begin{figure}

\includegraphics[scale=0.65]{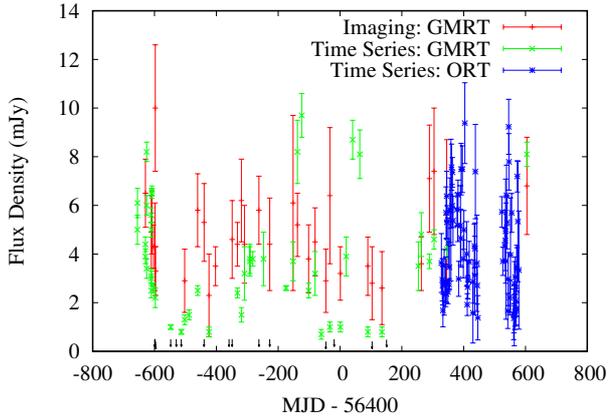}
\caption{Flux density measurements of PSR J1838+1523 at different epochs. The observing frequency is 325 MHz. Red crosses indicate flux density measured from GMRT images. Green crosses indicate flux density measured from GMRT time-series data and blue asterisks indicate flux density measured from ORT time-series data. Black arrows, near the bottom of the plot, indicate eight sigma upper limits on the flux density for a pulsed source in time-series non-detections.}
\label{1838_fluxvar}

\end{figure}

We made flux density measurements in the radio images using the AIPS task \textsc{jmfit}. We used a single elliptical Gaussian as a starting model to fit the observed source. The major and minor axes of the post-fit position error ellipse reported by \textsc{jmfit} were comparable to the synthesized beam size. Thus, this source is an unresolved (point) source and possesses no structure greater than 6$''$ in angular size. We have plotted the measured flux density as a function of MJD for imaging as well as time-series data in Figure \ref{1838_fluxvar}. As can be clearly seen in the plot, PSR J1838+1523 sometimes shows continuum radio emission, when it is not detected in pulsation.

The mean flux density of PSR J1838+1523, as estimated from all the measurements, at different frequencies is listed in Table \ref{1838_fluxes}.

\begin{table}
\begin{center}
\begin{tabular}{ccc}

\hline

Observation        & \multicolumn{2}{c}{Frequency}                 \\
type               & 325 MHz           & 610 MHz                   \\

\hline

Time-series (GMRT) & 4.0(1.8) mJy & 1.0(0.4) mJy         \\
Time-series (ORT)  & 4.2(1.6) mJy & -                         \\
Imaging (GMRT)     & 4.7(1.2) mJy & 1.4(0.7) mJy         \\

\hline

\end{tabular}
\caption{Mean flux densities of PSR J1838+1523 at different frequencies. The uncertainties (reported in parentheses) come from the systematic variation across time rather than measurement uncertainties, which are about 15\%.}
\label{1838_fluxes}
\end{center}
\end{table}

All the measurements at a given frequency are consistent with each other with the mean flux density of 4.3$\pm$1.8 mJy at 325 MHz and 1.2$\pm$0.7 mJy at 610 MHz. This implies a spectral index of $-$2$\pm$0.8. Extrapolating the spectral index to a frequency of 1400 MHz gives a predicted flux density of about 0.2 mJy which is much lower than the detection limit of 2.5 mJy for NVSS \citep{ccg+98}. The rms noise in the 1170 MHz maps made with the GMRT imaging data were of the same order, whereas the expected flux density at 1170 MHz is 0.3 mJy. Thus, the non-detection in the NVSS maps and 1170 MHz GMRT observations is consistent with the implied spectral index. In fact, given the level of uncertainties in the imaging measurements at 610 MHz (Table \ref{1838_fluxes}), the non-detections at this frequency are not surprising considering the fact that the flux density of the pulsar is very close to the detection limit and a small change due to scintillation will take the flux density below the detection limit.

\subsection{Flux density variations}
\label{fluxvar}

One look at Figure \ref{1838_fluxvar} reveals that the flux density of PSR J1838+1523 varies a lot. In order to estimate the scintillation parameters, we carried out a five hour observation with the ORT on 2015 June 4 (MJD 57177). The data were recorded using 1024 frequency channels and were sampled every 512 $\mu$s to achieve maximum time and frequency resolution. No clear scintillation features were visible in the dynamic spectra. For this line of sight, the predicted diffractive inter-stellar scintillation (DISS) time scale is 300 s, while the scintillation bandwidth is 0.25 MHz \citep{cl02}. The lack of clear scintillation features prevents us from drawing a firm conclusion about the exact cause of flux density variations, which could be due to refractive inter-stellar scintillation (RISS).

\section{Discussion}
\label{disc}
This survey detected 28 pulsars including one discovery. The simulations using \textsc{psrpop}\footnote{http://www.psrpop.phys.wvu.edu/index.php} \citep{lfl+06} with the post-analysis parameters (including the sensitivity degradation factor and reduced effective bandwidth due to RFI clipping) indicated the detection of 91 to 101 normal pulsars and a maximum of five millisecond pulsars (MSPs) over the full survey region. Out of these, 72 normal pulsars and three MSPs were already discovered in previous surveys [see the ATNF pulsar catalogue \citep{mht+05} at http://www.atnf.csiro.au/people/pulsar/psrcat/]. As can be seen in Figure \ref{obs_point}, we could only cover 10\% of the proposed survey region, that too with a very patchy sky coverage. To compare expected number of pulsars with the yield of our survey with this patchy coverage, we tried to obtain the detection rates for strips of 10\% area along Galactic latitude and longitude using \textsc{psrpop}. These simulations suggest detecting between four to 15 normal pulsars and between zero to two MSPs. Our detection of 28 pulsars, including two MSPs and one discovery, is therefore higher than that predicted by the population models, which were used in the simulations. While this may suggest revision of these models, it may also be noted that the number of known pulsars in the region observed by us is almost half of that in the full proposed region (30 normal pulsars and two MSPs). Thus, it is also possible that this distinctly non-uniform survey coverage in terms of the distribution of known normal pulsars may be responsible for the higher inferred detection rate.

Broad-band RFI also affected the yield of our survey. The analysis without implementing zero DM filtering resulted in the detection of only 10 pulsars in the periodicity search and four in the single-pulse search. After zero DM filtering, the degradation factor for the periodicity search changed from about 2.5 to 1.9 accompanied with the detection of 18 more pulsars including one discovery, while the change in degradation factor for the single-pulse search was from 2.5 to 1.3 with the detection of nine more pulsars. Thus, the only pulsar discovered in this survey was detected because of the zero DM filtering technique. Although broad-band RFI at low radio frequencies can not be completely removed by zero DM filtering, these figures demonstrate the usefulness of this technique and the need for doing more sophisticated RFI mitigation for low frequency pulsar surveys. 

The follow-up observations for PSR J1838+1523 and the subsequent analysis allowed us to obtain a timing solution in a step-wise approach using different types of measurements with multiple telescopes. The time-series observations using a gridding technique and initial timing observations with the GMRT provided the first constraints on the position. The position uncertainty in right ascension thus obtained was smaller than that in declination. The unique beam shape of ORT then allowed the uncertainty on the declination position to be reduced. Further refinement in position used detection with the GBT, which has a much narrower beam ($\sim$12$'$) at 820 MHz, detections using PA mode of GMRT at 325 and 610 MHz and multi-frequency imaging data with the GMRT. Apart from a much more refined pulsar position, the same observations provided further constraints on the flux density, spectral index and flux density variations completely characterizing the source under study. Rapid localization, important for a reliable timing solution, thus requires using multiple techniques and telescopes. Such a strategy will be vital for the Square Kilometre Array (SKA), which is expected to find more than 10,000 normal pulsars and 1,500 MSPs \citep{kbk+15}.

PSR J1838+1523 shows both short-term and long-term flux density variations (Figure \ref{1838_fluxvar}). While this is typical of most pulsars, the flux density of this pulsar varies by more than an order of magnitude, when the upper limits in Figure \ref{1838_fluxvar} are taken into account, which is unusual. This was the reason for a sequence of non-detections after its discovery, which presented a difficulty in its confirmation and localization. Given the small scintillation bandwidth, we believe that DISS may not be responsible for its non-detection, but could account for the flux density variation seen in epoch to epoch measurements. The variations on shorter time scales (like the ones seen in the ORT data with daily cadence) could be produced by DISS, whereas the non-detections in time-series data at GMRT in the longer term could be caused by long-term RISS given the clustering of GMRT time-series flux densities near the detection limit. 

The fact that we detect a continuum source even in the epochs of time-series non-detections hints that either the interferometric data are more sensitive than the IA data or we are seeing un-pulsed emission in the image, similar to what was seen for PSR J0218+4232 by \cite{ndf+95}. The flux density ratio of continuum to pulsed emission from their studies is of the order of 2. In their work, they recommend that the flux densities should be measured simultaneously to establish that the variation is indeed intrinsic. This is exactly the case for PSR J1838+1523. If we take the time-series non-detections into account, the ratio of continuum to pulsed emission could be as high as 3.5 for PSR J1838+1523. In fact, we did not see any appreciable pulsed emission even after coherently folding nine out of the 15 non-detections (the remaining data had a lot of RFI). If the pulsar does switch off completely, this would imply a ratio of ON to OFF state emission of about 23.5, much less than what was estimated for other intermittent pulsars \citep{klo+06,crc+12,llm+12}. We also do not see any evidence for a change in the period derivative of the pulsar unlike the intermittent pulsars. This may hint at the failure of the coherent emission mechanism responsible for pulsed emission while keeping the same amount of particle flow, which may have given rise to the apparently un-pulsed emission.

\section{Conclusions}
\label{concl}
We have carried out a blind pulsar survey with the GMRT at 325 MHz, resulting in the detection of 28 pulsars including one new discovery, PSR J1838+1523. The overall survey sensitivity for an eight sigma detection for 10\% duty cycle is estimated to be 2.7 mJy. The newly discovered pulsar, PSR J1838+1523, has a period of 549 ms and a DM of 68 pc cm\sups{$-$3}. We have obtained the full timing solution of this pulsar by using a combination of radio imaging (aiding the localization) and multi-frequency timing observations carried out with the GMRT, GBT, and ORT. We see a lot of variation of the flux density of this pulsar over short (days) and long (months) durations, including many non-detections. We believe that the short-term variations could be intrinsic and longer term variations could be due to refractive scintillation. We also see continuum radio emission from this pulsar at 325 MHz even when there are no pulsations detected. From our study of PSR J1838+1523, we conclude that simultaneous imaging capability is highly desirable for future low frequency pulsar surveys and note that pulsar surveys like the one mentioned here would provide good testing grounds for future pulsar surveys with the SKA. \\

\vspace{24pt} 

\noindent \textbf{ACKNOWLEDGEMENTS} \\

\vspace{1pt}
\noindent We thank the staff of the ORT, and the GMRT who have made these observations possible. The GMRT is run by the National Centre for Radio Astrophysics of the Tata Institute of Fundamental Research. We thank West Virginia University for its financial support of GBT operations, which enabled some of the observations for this project. MPS would like to thank Ishwara Chandra, and Ruta Kale for the help regarding advanced imaging techniques. We thank the referee for their comments, which have improved the manuscript. This work made use of PONDER receiver, which was funded by XII plan grant TIFR 12P0714. This work made use of the NCRA high performance cluster (HPC) facility, which was funded by XII plan grant TIFR 12P0711. We would like to thank V. Venkatasubramani, Lt. Shekhar Bachal and the whole HPC team. BCJ, MAK and PKM acknowledge support from DST-SERB Extra-mural grant EMR/2015/000515. MPS, MAM and DRL acknowledge support from NSF RII Track I award number OIA--1458952. MPS, MAM and DRL are members of the NANOGrav Physics Frontiers Center which is supported by NSF award 1430284.

\bibliography{ref}

\bsp	
\label{lastpage}

\end{document}